\documentclass[10pt]{article}

\usepackage{amsmath}
\usepackage{amssymb}

\usepackage{graphicx}

\usepackage{cite}

\usepackage{color} 

\usepackage[labelfont=bf,labelsep=period,justification=raggedright]{caption}

\makeatletter
\renewcommand{\@biblabel}[1]{\quad#1.}
\makeatother

\date{}

\begin{document}

\begin{flushleft}
{\Large
\textbf{Optimal vaccination in a stochastic epidemic model of two non-interacting populations}
}
\\
Edwin C. Yuan$^{1}$, David L. Alderson$^{2}$, Sean Stromberg$^{1}$,  and
Jean M. Carlson$^{1,\ast}$ 
\\
\bf{1} Physics Department, University of California, Santa Barbara, California, \\
\bf{2} Operations Research Department, Naval Postgraduate School, Monterey, California 
\\
$\ast$  Corresponding E-mail: carlson@physics.ucsb.edu
\end{flushleft}

\section*{Abstract}
Developing robust, quantitative methods to optimize resource allocations in response to epidemics has the potential to save lives and minimize health care costs. In this paper, we develop and apply a computationally efficient algorithm that enables us to calculate the complete probability distribution for the final epidemic size in a stochastic Susceptible-Infected-Recovered (SIR) model. Based on these results, we determine the optimal allocations of a limited quantity of vaccine between two non-interacting populations. We compare the stochastic solution to results obtained for the traditional, deterministic SIR model. For intermediate quantities of vaccine, the deterministic model is a poor estimate of the optimal strategy for the more realistic, stochastic case. 
 

\section*{Introduction}

As rapid, long-range transportation becomes increasingly accessible, transmission of infectious diseases is a growing global concern.  Advances in biomedical therapies and production have enabled the development of large quantities of pre-pandemic vaccine ~\cite{leroux}.  The United Kingdom, Japan, and the United States have plans to stockpile 3.3 million, 10 million, and 40 million doses, respectively, of pre-pandemic H5N1 vaccine~\cite{keeling2}.  However, in the face of a spreading pandemic, even seemingly extensive resources would be insufficient to provide global coverage, mandating the development of effective protocols for the allocation of limited vaccine~\cite{wu}~\cite{knipl}. 

A starting point for many studies of disease transmission in populations is the Susceptible-Infected-Recovered (SIR) model introduced by Kermack and McKendrick~\cite{kermack}. In this model, at any given time each individual is in one of three states. The dynamic evolution of the population is described by two irreversible transition probabilities: one describes the rate at which a susceptible individual becomes infected, and the other describes the rate at which an infected individual recovers (or dies).  The transition rates can be combined, and the SIR model can be parameterized by a single number, the reproductive number, $r_0$, which characterizes how effectively the infective agent moves through the population. On average, $r_0$ describes the number of susceptible individuals infected by a single infected individual in a population of susceptible individuals.  Reducing the number of susceptible individuals in a population, via vaccination for example, decreases the effective reproductive number $r_{\mathrm{eff}}$.  When $r_{\mathrm{eff}}<1$, the number of infected individuals tends to decline, whereas if $r_{\mathrm{eff}}>1$, the number tends to grow.

The concept of a reproductive number for an infectious disease can be generalized to more complex models of epidemiology. Previous work on developing optimal vaccination strategies typically focus on minimizing $r_{\mathrm{eff}}$, either by proactive dispersal of vaccine before the infection reaches a population~\cite{anderson}, or reactive dispersal~\cite{gaff}~\cite{klepac}, after infection has been detected in a group. This lowers the overall size of the epidemic, measured by the total number of individuals who have been infected throughout the course of the epidemic, by suppressing the rate of infections.  

Numerous computational studies of large-scale veterinary infections, such as foot-and-mouth disease~\cite{tildesley}~\cite{keeling1}, Johne's Disease~\cite{groenendaal}, as well as human infections like measles~\cite{grenfell} and SARS~\cite{hufnagel}, have been performed. In models that aim to capture field observations, detailed, case-specific information such as demographics of the population, timing and logistics for vaccine deployment, delays associated with the immune response, and overall vaccine efficacy are often essential to the investigation. In all cases, there are tradeoffs between complexity and realism, and between computational viability and the generality of the results.

In this paper, we abstract the geographic, demographic, and disease specific information, and instead focus on the fundamental problem of stochastic SIR dynamics with prophylactic vaccination in two non-interacting populations (e.g. two well-separated cities). Previously, Keeling and Shattock~\cite{keeling2} considered the deterministic SIR model in this scenario, and obtained striking results. As the total amount of available vaccine is increased, the allocation of vaccine that minimized the total number of infected individuals can undergo discontinuous transitions.  With a small amount of vaccine, the optimal strategy involves ensuring that the smaller population was well protected foremost.  However, with enough vaccine, the optimal strategy switches abruptly to protecting the large population, leaving the smaller population entirely unprotected. These results were well explained in terms of a phenomenon referred to as \lq\lq herd immunity," whereby immunization of a fraction of a population protects even those who are not vaccinated, by reducing the effective reproductive number $r_{\mathrm{eff}}$ to a value below unity. Vaccination removes susceptible individuals from the population. If there are less susceptible individuals in the population, on average, an infected individual will infect fewer individuals. Herd immunity occurs when $r_{\mathrm{eff}}<1$, i.e., on average, at the start of the epidemic each infected individual transmits his disease to less than one person.  Keeling and Shattock explained the sharp transitions in the optimal strategy as arising from a strategy that aims to induce herd immunity in the largest population possible.

While the deterministic SIR model is characterized by two coupled ordinary differential equations, the stochastic SIR model involves a high dimensional state space with probabilistic transitions between partitions of the overall population, characterized by the number of individuals in each state. Stochasticity leads to noteworthy differences in the epidemic size. When $r_{\mathrm{eff}}>1$ the probability distribution for the total epidemic size is bimodal~\cite{gordillo}, comprised of a roughly Gaussian peak centered at the deterministic epidemic size, as well as a second peak for small, \lq\lq terminal infections," describing the likelihood the disease will fail to propagate significantly during the initial phase of infection. The peaks of the distribution are well separated when the population size is large, so that if the number of infected individuals exceeds a critical size, the epidemic progresses to a large size, characterized on average by the deterministic results. However, the non-negligible probability that the disease will fail to propagate in a given population results in significant differences for optimal allocation of vaccine over a wide range of parameters. 

The rest of this paper is organized as follows. In Methods we review the deterministic SIR model, and its stochastic generalization. We approach the stochastic problem using a master equation for the time evolution of the complete probability distribution for the number of individuals in each state. Building on the computationally efficient algorithm recently developed by Jenkinson and Goustias~\cite{jenkinson}, we introduce a modification which leads to even greater computational efficiency. In Results we compute the probability distribution for the final epidemic size for a range of parameters to identify regimes for which the stochastic and deterministic models differ most significantly. We compute the optimal allocation of vaccine between two non-interacting populations, and compare our results with the deterministic case. Stochastic effects are most pronounced in situations involving an intermediate amount of resource availability. We conclude with a discussion of our results and future directions.

\newpage
\section*{Methods}

We briefly review the deterministic SIR model~\cite{kermack}, a system of coupled differential equations for modeling the growth of an epidemic in a well-mixed population within which all agents interact equally with all other agents.  The model describes a population of $N$ individuals divided into three classes, susceptible \textbf{S},  infected \textbf{I}, and recovered \textbf{R}:
\begin{align}
                    		\frac{d\textbf{S}(t)}{dt}&=-\beta \textbf{S}(t)\textbf{I}(t);   \\
			\frac{d\textbf{I}(t)}{dt}&=\beta \textbf{S}(t)\textbf{I}(t)-\gamma \textbf{I}(t); \\
			\frac{d\textbf{R}(t)}{dt}&=\gamma \textbf{I}(t).
                  \end{align}
  We may omit the equation for the recovered class because we can always deduce the number of recovered individuals from the fact that the total number individuals in the population, $N$, is fixed, so that $\textbf{R}(t)$=$N$-\textbf{S}(t)-\textbf{I}(t). 

Equations 1-3 can be thought of as a mean field theory where the continuous variables $\textbf{S}(t)$ and $\textbf{I}(t)$ are the average values (over many iterations) of two discrete integer-valued variables $S$ and $I$.  At any time then, the system can be characterized as being in a state $(S,I)$, which can undergo one of two transitions:
\begin{eqnarray*}
\text{}(S,I)\rightarrow( S-1,I+1)\text{ at rate }\text{}\beta SI;   \nonumber \\
\text{}(S,I)\rightarrow( S,I-1)\text{ at rate }\text{}\gamma I.  \nonumber
\end{eqnarray*}
 
\noindent The parameters $\beta$ and $\gamma$ can be defined in terms of physical observables, the average number of contacts each person makes per day $c$, the probability of infection through contact $p$, and the characteristic duration of the infection $T$:
\begin{gather}
\beta=
\frac{
  \begin{pmatrix}
    \text{rate at which each}\\
    \text{individual makes contacts}
  \end{pmatrix}
  \times
  \begin{pmatrix}
    \text{probability of infection}\\
    \text{from contact}
  \end{pmatrix}
}{
  \begin{pmatrix}
    \text{total population size $N$} 
  \end{pmatrix}
}
=\frac{c\times p}{N};
\end{gather}
\begin{gather}
\gamma=
\frac{
  1
}{
   \text{characteristic duration of infection} 
}
=\frac{1}{T}.
\end{gather}

For each set of parameters $\beta$ and $\gamma$, the reproductive number, $r_0$ is defined to be:
\begin{equation}
r_0=\frac{\beta \textbf{S}_0}{\gamma}= c\times p\times T \times \frac{ \textbf{S}_0}{N},
\end{equation}
where $\textbf{S}_0$ is the initial number of susceptible individuals, $\textbf{S}(t=0)$.  Here $r_0$ can be interpreted as the average number of new infections a single infected individual will produce in a completely susceptible population.  Thus if $r_0<1$, in the deterministic model $d\textbf{I}/dt<0$, the number of infected individuals will decline from the initial seed value, $\textbf{I}(t) \leq \textbf{I}_0=\textbf{I}(t=0)$, and no epidemic will occur.  In our numerical simulations, the value of $r_0$ is tuned by varying $\beta$.

In this paper we investigate the effects of prophylactic vaccination.  Vaccinating $V$ individuals proactively corresponds to removing to $V$ susceptible individuals before the epidemic begins, thus lowering the effective reproductive number.  This assumes that the vaccine is completely effective.
Let $r_0$ denote the reproductive number prior to vaccination, and $r_{\mathrm{eff}}$ denote the effective reproductive number which is achieved after vaccinating $V$ individuals:
\begin{eqnarray}
r_{\mathrm{eff}}= c \times p \times T\times \frac{ \textbf{S}_0-V}{N}
           =r_0(\frac{\textbf{S}_0-V}{\textbf{S}_0}).
\end{eqnarray} 
If a sufficient number of individuals $V$ are vaccinated, $r_{\mathrm{eff}}$ may be reduced to a value below unity, so that $d\textbf{I}/dt<0$, and, as a result, the epidemic will not grow.  Thus the entire population will be safeguarded without vaccinating the entire population.  This phenomenon is known as herd immunity.

\vspace{1cm}
In the stochastic SIR model, the infection and recovery reactions are modeled as continuous-time Markovian processes.  Let $\phi_{S,I}(t)$ be the probability at time $t$ of a population with $S$ susceptible individuals and $I$ infected individuals with $N=S_0+I_0$ where $S_0$ and $I_0$ are the initial values of $S$ and $I$.  The evolution of $\phi_{S,I}(t)$ in time is then governed by:
\begin{align}
                		\frac{d}{dt}\phi_{S,I}(t)=&\beta (S+1)(I-1) \times\phi_{S+1,I-1}(t) \nonumber \\
			&+\gamma (I+1) \times \phi_{S,I+1}(t) \nonumber \\
                                 &- (\beta SI+\gamma I )\times\phi_{S,I}(t),
 		\end{align}
where the first two terms on the right hand side correspond to transitions into the state $(S,I)$ by a susceptible individual becoming infected or an infected individual recovering, respectively, and the third term corresponds to the probability of leaving the state $(S,I)$ through infection or recovery.  While the deterministic model tracks the time evolution of two ensemble averaged variables $\textbf{S}(t)$ and $\textbf{I}(t)$, the stochastic model has up to $(S_0+I_0+1)(S_0+1) \sim N^2$ possible states ($I$ can take values $0$ to $S_0+I_0$ and $S$ can take values $0$ to $S_0$).  All the probabilities $\phi_{S,I}(t)$ can be assembled into a vector $\vec{\phi}$ and the entire system of equations can then be written in matrix form.  

We computationally integrate this system of equations using a modified version of Jenkinson and Goutsias's method~\cite{jenkinson} of Implicit-Euler integration.  The matrix $\mathbb{A}$ consists of the coefficients of Equation 8, and describes the transition probabilities:

\begin{eqnarray}
d{\vec{\phi}}&=&\mathbb{A}\vec{\phi} dt.
\end{eqnarray}
The above equation is discretized by introducing a time step, which controls the accuracy of the method:
\begin{eqnarray}
\vec{\phi}_{t_{i+1}}=&\vec{\phi}_{t_i}+\mathbb{A}\vec{\phi}_{t_{i+1}}(t_{i+1}-t_i); \nonumber \\
=&(\mathbb{I}-\mathbb{A}\Delta t)^{-1}\vec{\phi}_{t_i}.
\end{eqnarray}

\noindent For computational efficiency, it is essential to order the components of the vector $\vec{\phi}$ in such a way so that $\mathbb{A}$ is lower triangular. We made modifications to the way the algorithm counts states, enabling considerably faster computational speed, especially as the population size increases.  Where Jenkinson and Goutsias take the approach of counting the so-called ``degree of advancement," a scenario in which each state corresponds to a specific sequence of reactions, we instead take the ``population process" approach by enumerating all states of the system without tracking which reactions might have led the system to the state in question.  In both methods one begins with $(S_0+I_0+1)(S_0+1)$ states.  In our method, we remove those states that have zero probability of occurring, but are naturally included in the degree of advancement procedure.  For example, many states where $S+I> N$ are retained in the degree of advancement procedure but are explicitly excluded in our method.  The result is that we track $[(S_0+1)(I_0+1)+(S_0+1)(S_0)/2]$ states, which in the limit $I_0 \ll S_0$ is approximately $\sim N^2/2$.  As the system size $N$ grows, the difference in the total number of states between the two methods can significantly impact the time it takes to integrate the system of equations.

\newpage
\begin{figure}[!htb]
\center
\includegraphics[width=\linewidth,draft=false]{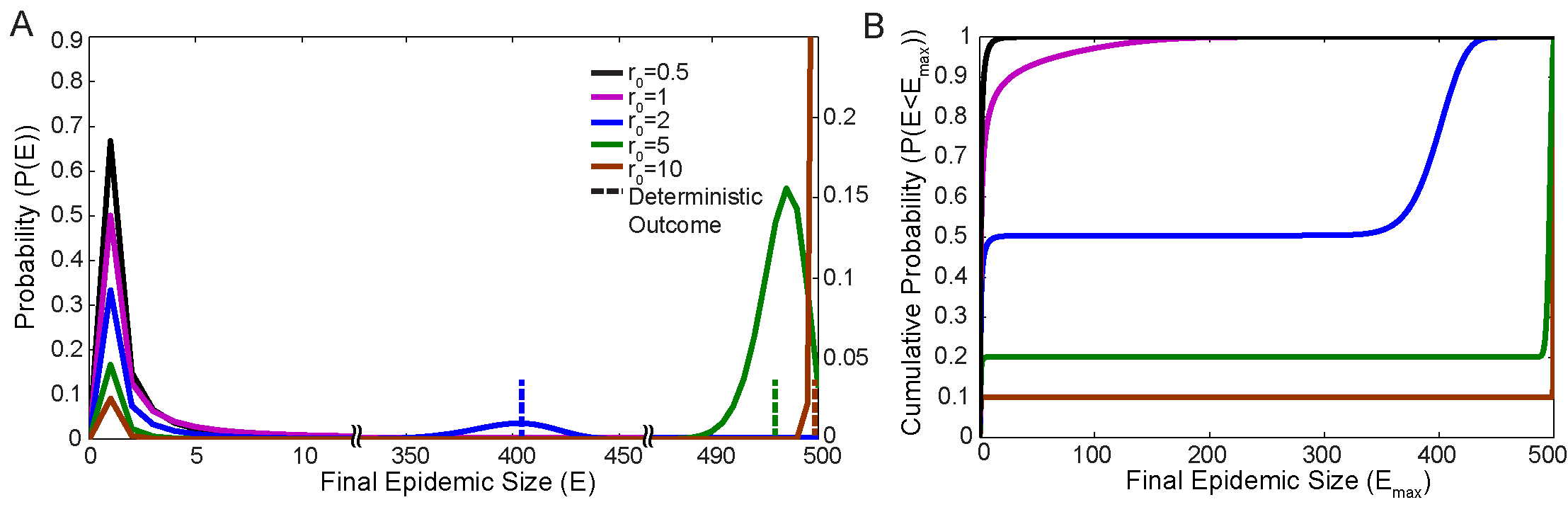}
\caption{\textbf{Epidemic size distribution:}  Figure A illustrates the final epidemic size distributions $P(E)$ for various values of $r_0$ in the stochastic SIR model.  The corresponding result for the deterministic SIR model (a number \textbf{E}) is marked in each case  by a dashed line.  The left vertical scale  in Figure A describes the small terminal infections (left of the first horizontal scale break), and the right vertical scale describes large-scale infections (right of the first horizontal scale break). Figure B illustrates the corresponding cumulative probability distribution $P(E<E_{\mathrm{max}})$ or the probability of an epidemic of size less than $E_{\mathrm{max}}$.  This also shows the relative weight in the terminal infection and in the large-scale epidemic.  In each case $N=500$ and $I_0=1.$}
\end{figure}

All numerical calculations are performed using MATLAB. The system is initialized with a population size $N$, $I_0$ infected individuals, and an initial reproductive number $r_0$.  Thus, at time $t=0$ the probability of state $\phi_{N-I_0,I_0}(0)=1$ and the probability of all other states equals zero.  The collection of probabilities $\vec{\phi}$ of all accessible states is then evolved forward in time until the distribution reaches a stationary state where the probability of having any state $(S,I)$ where $I>0$ is vanishingly small. At that point, all individuals in the initial population of size $N$, have either been infected, and are now recovered, or remain susceptible. For the parameters considered here, we observe that an integration time of $t=200$ is sufficient in all cases. Once the simulation is complete, we define the final epidemic size $E$ as: 
\begin{align}
E&=\lim_{t \to \infty} N-S(t) \text{ in the stochastic model};\nonumber \\
\textbf{E}&=\lim_{t \to \infty} N-\textbf{S}(t)  \text{ in the deterministic model}.
\end{align}

Figure 1 illustrates numerical results for the epidemic size distribution $P(E)$, describing the probability of having a total of $E$ individuals infected over the course of the entire simulation period.   We observe that for $r_0>1$ the probability distribution consists of two parts.  On the left side of Figure 1A, there is peak describing small, \lq\lq terminal infections," which fail to propagate significantly in the population (i.e. the infection terminates before a large number of individuals are impacted). The peak describing terminal infections decays approximately exponentially from the peak value at $P(I_0)$. 
In the stochastic model, there is always a nonzero probability the infection will end without becoming a large scale epidemic. 

When $r_0>1$, the distribution $P(E)$ exhibits a second peak towards the right side of Figure 1A, describing  \lq\lq large-scale epidemics." This peak is approximately centered at the epidemic size predicted by the deterministic SIR model, illustrated for each value of $r_0$ by the corresponding vertical dashed line in Figure 1A. The size of the large-scale epidemic scales with the size of the population $N$, resulting in a larger separation of the peaks for increasing population sizes. To quantify the likelihood of a terminal infection versus a large-scale epidemic by the relative weight associated with each of the peaks, numerically we define the point separating the terminal infection and the large-scale epidemic to correspond to the local minimum in probability that exists between the two peaks. The likelihood of a terminal infection, represented by the total weight in the terminal infection peak, decreases with increasing values of $r_0$ and $I_0$. As $r_0$ approaches unity from above, the large-scale epidemic progressively decreases in mean size, but increases in variance. Eventually the distinction between terminal infections and large-scale epidemics vanishes (the local minimum in $P(E)$ ceases to exist). This is associated with a critical phase transition~\cite{lalley}, and occurs at a value of $r_0$ that approaches unity as the population size $N$ tends to infinity. 
When $r_0\le 1$, the probability distribution is described only by terminal infections.  

The cumulative epidemic size distribution $P(E<E_{\mathrm{max}})$ describes the probability of having an epidemic of size less than $E_{\mathrm{max}}$, and is shown in Figure 1B.  The extended flat portions of the curves indicate that a population of $N=500$ individuals is well within the large population limit, defined by a large separation between the terminal infection and large-scale epidemic, with little probability of observing an epidemic size in between the two.  Figure 1B also illustrates how the total probability is distributed between the terminal infection and the large-scale epidemic.  The smaller the value of $r_0$, the greater the likelihood that the initial seed population of infected individuals will fail to spread the disease.

\newpage
\section*{Results}
 
Our aim is to highlight key differences between stochastic and deterministic approaches to developing a framework for the optimal allocation of vaccine between two non-interacting populations. We begin by observing how a single population reacts to different levels of vaccination in the stochastic and deterministic SIR models; the results provide the basis for the optimization process.  Subsequently, we determine the optimal deterministic and stochastic solutions that minimize the average epidemic size, and contrast their properties.  Finally, we consider an alternative optimization based on imposing a maximum tolerance for the epidemic size and show that the stochastic optimal solution better fulfills this measure than the deterministic optimum.

\subsection*{Impact of Vaccination on the Epidemic Size of a Single Population}

We first consider how the epidemic size within a single population decreases as a function of increasing vaccine allocation.  Vaccine allocation $V$ removes $V$ susceptible individuals from the initial state $(S_0,I_0)\rightarrow(S_0-V,I_0)$ after which the stochastic SIR model evolves according to Equation 8 (Equations 1-3 for the deterministic SIR model).  The resulting dynamics determine the size of the epidemic according to Equation 11. Decreasing the initial number of susceptible individuals $S_0$ by $V$ will not in general lead to a corresponding reduction $V$ in the final epidemic size. An important quantity for optimizing the allocation is the incremental reduction in the expected epidemic size per incremental increase in the allocation. We define this below as the \lq\lq gain" $G$.

In Figure 2 we illustrate the numerical results for a population of $N=500$ individuals with different numbers of initial infected individuals $I_0$ and different reproductive numbers $r_0$.  An amount of vaccine $V$ ($0\le V \le N-I_0$) is given to the population and we compute the average final epidemic size $\langle E \rangle$=$\int{P(E)\times E \,dE}$ as a function of $V$, where $P(E)$ is computed as in Figure 1A.  We also plot the corresponding deterministic curve in each case, where $P(\textbf{E})$ here is described by a $\delta$-function at the deterministic epidemic size $\delta(\textbf{E})$.  In the stochastic model, the quantity $\langle E \rangle$ depends on the statistics of both the terminal infection and the large-scale epidemic; $\langle E \rangle$ may not correspond to an epidemic size that is likely to be observed because there may be a large separation between the observed sizes of terminal infections and large-scale epidemics, with the mean size lying somewhere in between.
\
\begin{figure}[!htb]
\center
 \includegraphics[width=\linewidth,draft=false]{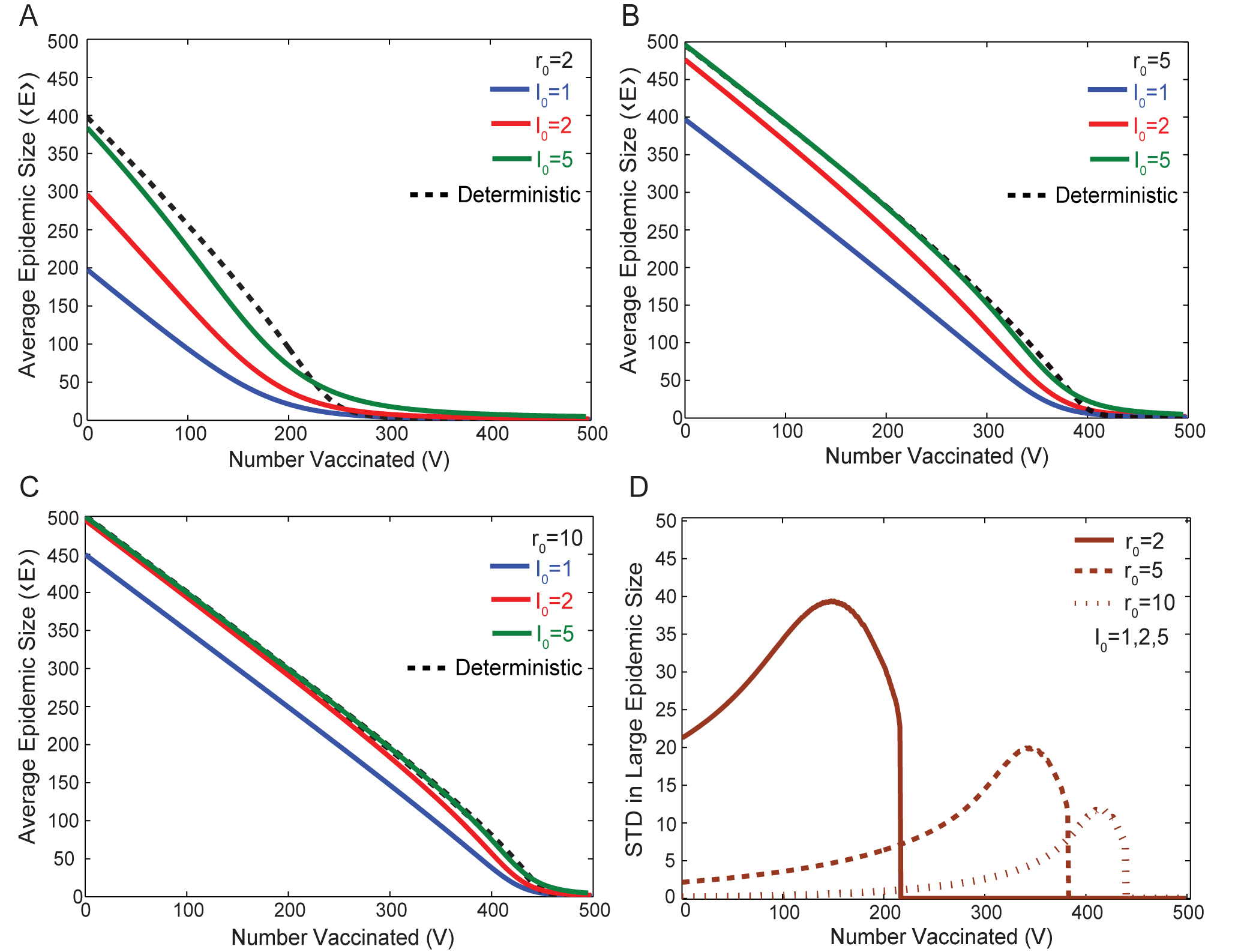}
\caption{\textbf{Comparison of the effects of vaccination between stochastic and deterministic models for various values of $r_0$ and $I_0$:}  Figures A, B, and C illustrate the  epidemic size as a function of the number of individuals vaccinated $V$.  For each value of $r_0$, the solid lines shows the average epidemic size $\langle E \rangle$=$\int{P(E)\times E \,dE}$, and dashed lines represent the corresponding deterministic epidemic size \textbf{E}.  The deterministic solution does not depend significantly on $I_0$ and is computed with $I_0=1$.  Figure D shows the standard deviation in the large-scale epidemic for various values of $r_0$ and $I_0$.  Theses curves are identical for all values of $I_0$ so only one such curve is shown. The value of $V$ defines the effective herd immunity threshold for each value of $r_0$. The results obtained here are for a population of $N=500$.}
\end{figure}

Herd immunity occurs in the deterministic SIR model when the initial effective growth rate of the number of infected individuals in the population becomes less than unity $(r_{\mathrm{eff}}<1)$~\cite{jong}, and is achieved at a value of $V$ determined by Equation  7, i.e. when $V/{\textbf{S}_0}=V/(N-\textbf{I}_0)=1-{r_0}^{-1}$. In the limit of large populations, the fraction that must be vaccinated to achieve herd immunity approaches $1-{r_0}^{-1}$. Thus for a population $N=500$, herd immunity occurs approximately when $V=250$ for $r_0=2$, $V=400$ for $r_0=5$, and $V=450$ for $r_0=10$.    
Approaching this value, the incremental reduction in expected epidemic size per increase in vaccine allocation increases monotonically. Note that the peak epidemic reduction rate occurs for a slightly smaller $V$ when $N$ is finite, compared to the $N\to\infty$ limit. This is due to the contribution of the initial seed population of infected individuals $\textbf{I}_0$ in the definition of $r_{\mathrm{eff}}$ and on the final size of the epidemic.

In the stochastic model, the corresponding transition is subtler. Increasing the vaccine allocation has three effects on $P(E)$. It decreases the mean size $\langle E\rangle$ and increases the variance of the large-scale epidemic, and also increases the relative likelihood of terminal infections. We associate the onset of \lq\lq effective herd immunity" in the stochastic model with the value of $V$ for which the distinction between terminal infections and large-scale epidemics ceases to exist, as measured by the existence of a local minimum in $P(E)$. Because of the probability of terminal infections, this generally occurs at a value of $V$ which is smaller than that of the herd immunity transition in the deterministic model. Furthermore, unlike the deterministic model, in the stochastic case approaching the onset of effective herd immunity does not coincide with a specific value of $r_{\mathrm{eff}}$ and is not generally the point of maximum impact per vaccine in the allocation (as measured by reduction in the average epidemic size).

Before the population has reached the deterministic herd immunity transition, i.e. when $r_{\mathrm{eff}}>1$, the deterministic epidemic size \textbf{E} (dashed lines of Figure 2A, B, C) is generally larger than the average stochastic epidemic size $\langle E \rangle$ (solid lines).  While the maximum size of the large-scale epidemic can be greater than the deterministic epidemic \textbf{E}, the average epidemic size $\langle E \rangle$ is smaller due to the fact that the stochastic model includes the possibility of a terminal infection. 

When sufficient vaccine is available to establish herd immunity in the deterministic model, the situation is reversed and the deterministic size \textbf{E} is generally smaller than the average stochastic epidemic size $\langle E\rangle$.  When $r_{\mathrm{eff}}<1$, in the deterministic model $d\textbf{I}/dt<0$ at t$=0$, and the initial number of infected individuals decreases.  On the other hand, stochastically there is always a possibility that the initial number of infected individuals will grow.  Hence, after the herd immunity threshold, the deterministic epidemic $\textbf{E}$ is smaller than the stochastic average epidemic $\langle E \rangle$.

The size of the average stochastic epidemic $\langle E \rangle$ also approaches the deterministic outcome \textbf{E} as both $I_0$ and $r_0$ become large.  A larger value of $r_0$ causes each infected individual to infect more susceptible individuals, while larger values of $I_0$ makes it less likely for every member of the initial group of infected individuals to recover before spreading their disease.  Both of these effects decrease the probability of a terminal infection. 

The standard deviation of the large-scale epidemic, which represents the fluctuation around the deterministic size of the epidemic \textbf{E}, is shown in Figure 2D.  It is independent of the initial number of infected individuals (i.e.~$I_0=1,2,5$) because statistics for large-scale epidemics are conditioned only on this portion of the distribution, which assumes the infection has progressed beyond a terminal infection, rendering the size of the initial seed population $I_0$ irrelevant, as long as it is not comparable to the size of the large-scale epidemic. The standard deviation does, however, increase with decreasing $r_0$, because a smaller value of $r_0$ implies each infected individual infects fewer susceptible individuals, and the epidemic is more likely to end before reaching the size \textbf{E} predicted by the deterministic SIR equations.

This standard deviation also is a function of $V$, the amount of vaccine allocated to the population.  It increases when $V$ is small, which can be attributed to the fact that having more vaccinated individuals reduces $r_{\mathrm{eff}}$.  This effect is balanced by the fact that more vaccinated individuals results in fewer available configurations ($S$,$I$) for the system to transition into.  Eventually the standard deviation peaks and then drops to zero, corresponding to the disappearance of the local minimum in $P(E)$ separating terminal infections from large scale epidemics, coinciding with our definition of effective herd immunity. The peak variance observed for $V$ slightly less than this value indicates that the largest uncertainty in the size of the large-scale epidemic is expected for allocations just below the effective herd immunity threshold. 

\newpage
An important quantity in determining the optimal allocation of vaccine is the ``gain" $G$, defined as the decrease in the epidemic size per initial susceptible individual removed by vaccination:
\begin{align}
G&=-\frac{d\langle E \rangle}{dV}\ \text{in the stochastic model}; \nonumber \\
\textbf{G}&=-\frac{d \textbf{E}}{dV}\ \text{in the deterministic model}.
\end{align}
 In the stochastic model, for small $r_0$, but greater than unity ($1<r_0\lesssim2.5$, for the other, fixed parameters considered here), the gain $G$ declines continuously from a maximum value at $V=1$.  This is illustrated explicitly for $r_0=2$ in Figure 3B.  This behavior implies that the larger the vaccine allocation given to the population, the smaller the benefits of additional vaccine.  For larger values of $r_0$ ($r_0\gtrsim2.5$) the gain instead initially increases (at a smaller rate than the corresponding deterministic curve), peaks, and then declines to zero.  In this case then, when deciding on vaccine allocation, there is a value of $V$ prior to reaching herd immunity where the gain from vaccination peaks.  This is shown for $r_0=5$ in Figure 3A.

\begin{figure}[!htb]
\centering
\includegraphics[width=\linewidth,draft=false]{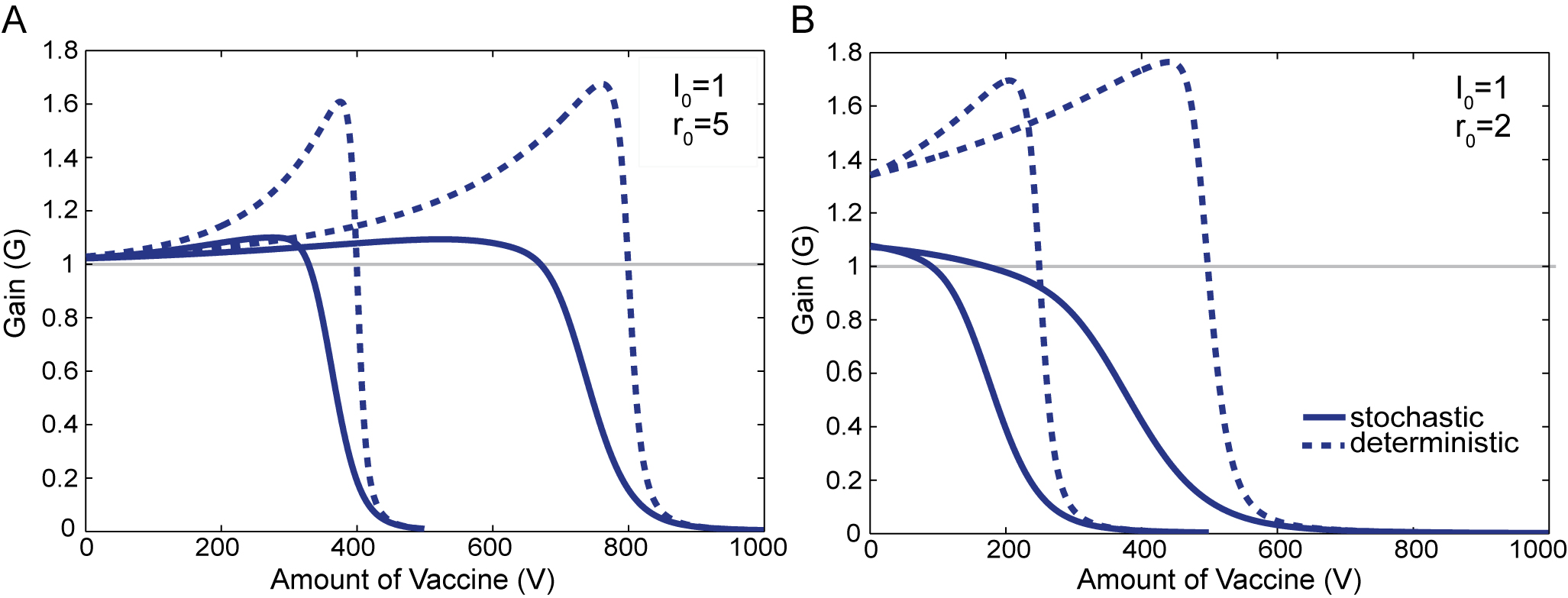}
\caption{\textbf{Gain $G$ as a function of the vaccine allocation $V$:}  For both the stochastic (solid lines) and deterministic (dashed lines) models, Figures 3A and 3B illustrate the gain $G$ (i.e. slopes) of the $I_0=1$ curves in Figures 2A and 2B, respectively. Figure A shows the gain $G$ for $r_0=5$ for both a population of $N^1=500$ individuals and also one of $N^2=1000$ individuals. Figure B shows the corresponding results for $r_0=2.$}
\end{figure}
                                                         
For all $r_0>1$ in the deterministic model, the gain curve follows the same qualitative pattern as the large $r_0$ case in the stochastic model. Initially, the gain $\textbf{G}$ increases, rising sharply prior to herd immunity, and then falling sharply after the vaccine exceeds the herd immunity point.  This means there is a significant increase in the gain $\textbf{G}$ from vaccination as the level of vaccine in the population nears the herd immunity threshold.

What is notably different between the stochastic and deterministic models is the sharpness of the peak and the rate of decline that follows.  This is apparent when comparing the stochastic and deterministic curves of Figure 3A.  This illustrates that beyond a threshold level of vaccination ($r_{\mathrm{eff}}<1$ in the deterministic model) there is almost no reduction in \textbf{E} by further application of vaccine to the population. In the stochastic model there is not as definitive a threshold level of vaccination.  The gain $G$ in the stochastic model begins to decline well before effective herd immunity is reached.  Thus in the stochastic model, the point of diminishing returns from vaccination will generally take place at smaller vaccine allocations $V$ compared to the deterministic model.   


%

\newpage
\subsection*{Optimal Vaccination Allocation for Two Populations}

Next we consider the problem of vaccine allocations for two non-interacting populations (e.g., two well-separated cities).  This scenario isolates a fundamental tradeoff in resource management, whereby allocating vaccine to one population occurs at the expense of the other.

Unless otherwise specified, we identify properties specific to each population with superscripts 1 and 2.  We assume in this section one population is relatively small ($N^1=500$ individuals), and the other is relatively large ($N^2=1000$ individuals). Both populations are initiated with a single infected individual $I_0^1=I_0^2=1$. A fixed total amount of vaccine $V$ ($0\le V \le1498$, where the maximum value of $V$ is given by $N^1+N^2-I_0^1-I_0^2=1498,$ accounting for one seed infected individual in each population) can be distributed between the two populations, so that the small population receives $V^1$, and the large population receives $V^2=V-V^1$. We define the optimal allocation to be the partition of $V$ into $[V^1,V^2]=[V^1,V-V^1]$ that minimizes the average total final epidemic size $\langle E \rangle$=$\langle E^1 \rangle$+$\langle E^2 \rangle$, where $0 < E\le N^1+N^2=N=1500$. In this scenario, the cost of producing and distributing vaccine is not taken into account, so it is always beneficial to use all of the available vaccine. 

Our objective is to determine the optimal solution as a function of $V$ for the stochastic SIR model. We compare our results to the corresponding optimal solution for the deterministic SIR model, which we also compute. This scenario was considered previously for the deterministic case in the limit of large population sizes ($N^1=100,000$ and $N^2=200,000$, using our notation) by Keeling and Shattock~\cite{keeling2}, who found that for a wide range of values of the reproductive number $r_0$, the optimal solution as a function of increasing $V$ was governed by the ability to induce herd immunity in the smaller population (small $V$), then in the larger population (intermediate $V$), and finally in both (large $V$). 
\begin{figure}[!htb]
\centering
\includegraphics[width=\linewidth,draft=false]{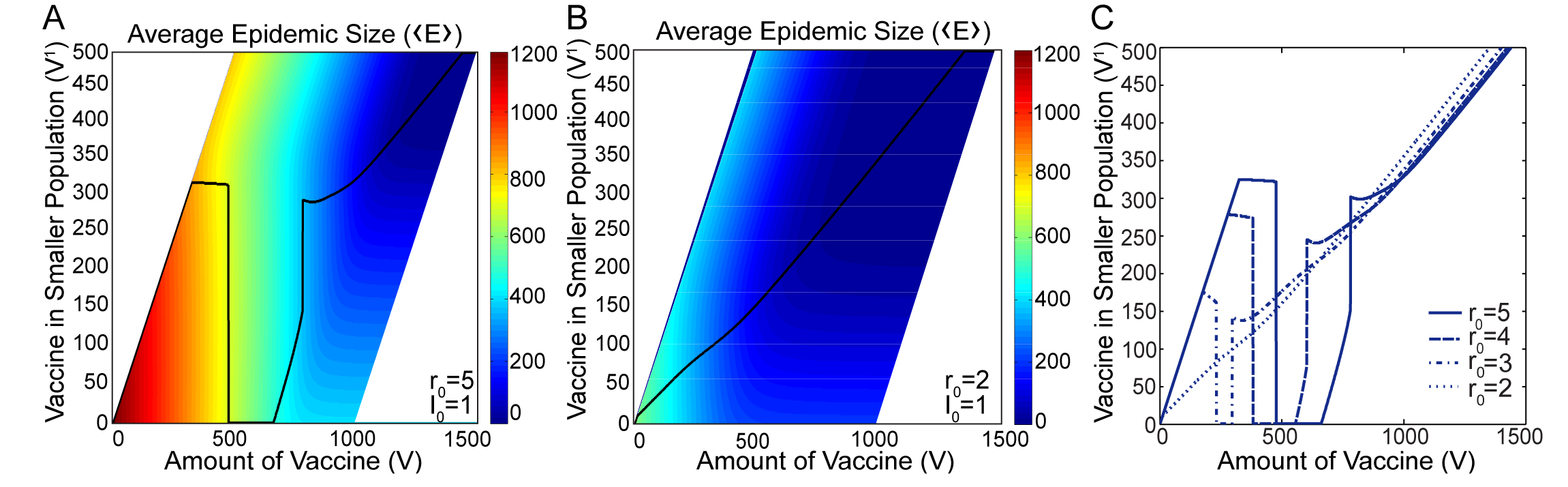}
\caption{\textbf{Optimal Solution in the Stochastic Model:}  The black lines of Figures A and B illustrate the optimal allocation of an amount of vaccine $V$ that minimizes $\langle E \rangle$ for, respectively, $r_0=5$ and $r_0=2$. The y-axis here shows the amount of vaccine allocated to the smaller population $V^1$.  The color scale indicates the average epidemic size $\langle E \rangle$ corresponding to a particular allocation of vaccine.   Figure C shows how the optimal solution varies as $r_0$ changes from 2 to 5. Switching behavior vanishes when $r_0\approx 2.9$ (not shown). Results are obtained for two populations of sizes  $N^1=500$ and $N^2=1000$. Both populations are initiated with a single infected individual $I_0^1=I_0^2=1$.}
\end{figure}

The colormap of Figures 4A and 4B illustrates the average epidemic size $\langle E \rangle$ corresponding to a particular allocation of vaccine, quantified in the figure  by the amount of vaccine in the smaller population $V^1$.  The optimal vaccine allocation is illustrated by the black line, along which $\langle E \rangle$ is minimized. The range in $V^1$ is limited by the constraints $V_1\le V-I_{0}^1=V-1$, and  $V_1\ge V-999$, i.e. neither population receives more vaccine than the number of initial susceptible individuals in that population.  This results in the limiting diagonals in the colormap of Figures 4A and 4B.


\vspace{.35cm}
\noindent \textbf{Switching:}
In the deterministic model, Keeling and Shattock\cite{keeling2} found that the optimal solution exhibited ``switching" behavior, in which the optimal vaccine allocation makes a significant, discontinuous, change when the total amount of vaccine $V$ exceeds a threshold size.  When the amount of vaccine $V$ is below this threshold size, the majority of the vaccine is optimally allocated to the smaller population. When the amount of vaccine $V$ is above this size, all of it is optimally allocated to the larger population.  This behavior persists for a wide range of reproductive numbers $r_0>1$ in the deterministic SIR model. 

The stochastic optimal solution exhibits switching behavior only for larger values of $r_0$ ($r_0\gtrsim 2.9$, for the other, fixed parameters considered here).  This is demonstrated for $r_0=5$ in Figure 4A.  Switching occurs first at $V=474$ and then again at $V=780$.   The first switching point, above which all vaccine is optimally allocated to the larger population, is present in both the stochastic and deterministic models, although the switching point of the stochastic model occurs at a smaller amount of vaccine $V$.  The second switching point is absent in the deterministic model.   The presence of the second switch in the stochastic model is explained in terms of the relative heights of the peaks of the gain curves in the next subsection.

For smaller $r_0$ ($1<r_0\lesssim 2.9$) in the stochastic model, there is no switching behavior, which is in contrast to the results of the deterministic model.  It is instead optimal to distribute any given total amount of vaccine $V$ approximately in proportion to the sizes of the populations themselves. This is shown for $r_0=2$ in Figure 4B.  The continuous transition between large $r_0$ values where switching does take place and small $r_0$ values where it does not is illustrated in Figure 4C.  As $r_0$ is decreased, the region between the two instances of switching behavior, where all vaccine is taken out of the smaller population and $V^1$=0, becomes narrower and disappears completely between $r_0=2$ and $r_0=3$ (at $r_0\approx 2.9$).  The difference between the $r_0$ dependence of switching in the stochastic and deterministic models is related to the presence of peaks ($r_0\gtrsim2.5$) in the gain curves.  This will be discussed in the following subsection. 

The conclusion for the stochastic optimal solution is that for small values of $r_0$ ($1<r_0\lesssim 2.9$), the optimal solution is to approximately distribute vaccine in proportion to population size.  For large $r_0$ ($r_0\gtrsim2.9$), two switches take place.  In the deterministic optimal solution, a single switch takes place for all values of $r_0>1$.

\vspace{.35cm}
\noindent \textbf{Understanding the Optimal Stochastic Solution:}
Next we examine the optimal stochastic solutions of Figure 4A and 4B in closer detail.  Figures 5A and 5B show the optimal stochastic solution as a solid line and the optimal deterministic solution as a dashed line, with both solutions represented by the fraction of the amount of vaccine $V^1/V$ given to the smaller population.  We seek to characterize the different strategies employed by the optimal solution in the different resource regimes, and also to quantify why the allocation transitions abruptly from one population to another.  More broadly, we explain why the optimal stochastic solution, which minimizes the average epidemic size $\langle E \rangle$, differs from the optimal deterministic solution, which minimizes the characteristic epidemic size \textbf{E}.   

\begin{figure}[!htb]
\centering
\includegraphics[width=\linewidth,draft=false]{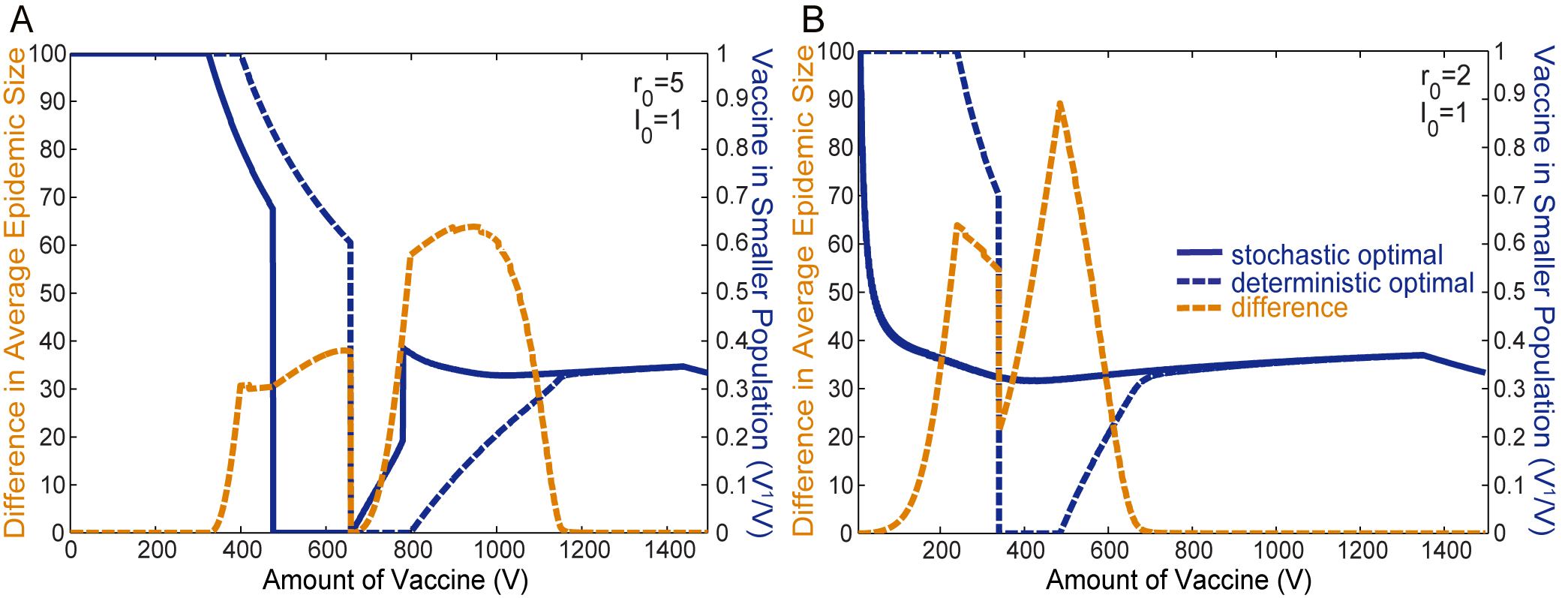}
\caption{\textbf{Optimal Protocols Represented as a Fraction of Total Vaccine:}  The solid blue line shows the optimal fraction of the available vaccine allocated to the smaller population $V^1/V$ in order to minimize the average stochastic epidemic size $\langle E \rangle$.  The dashed blue line shows a different optimal allocation that minimizes the deterministic epidemic size \textbf{E}. The right vertical scale applies to these two measurements. The difference in the resultant average epidemic size $\langle E \rangle$ between the two protocols when applied to the stochastic model is plotted in gold (measured by the left vertical scale).  The results are obtained for $r_0=5$ in Figure A, and $r_0=2$ in Figure B, and populations of sizes $N^1=500$ individuals and $N^2=1000$ individuals, both initiated with a single infected individual $I_0^1=I_0^2=1$.  }
\end{figure}

Much of our insight comes from analysis of the gain curves, which are shown in Figure 3A and 3B for $r_0=5$ and $r_0=2$.  By Equation 6, the area under each gain curve $G(V)$ up to a particular value of $V$ is the decrease in the epidemic size due to that amount of vaccine $V$.  We begin with the case when $r_0=5$, which is representative of the large $r_0$ regime ($r_0\gtrsim 2.9$) where switching does take place.

Figure 5A shows that with small amounts of vaccine $V$, all of the vaccine is optimally allocated to the smaller population. In the deterministic case, this strategy persists for larger amounts of vaccine $V$, unless there is enough vaccine for the small population to achieve herd immunity, at $V=400$. For a range of $V$ greater than $V=400$, herd immunity is preserved in the small population, and the remaining vaccine is optimally allocated to the large population.   For the stochastic case, vaccine allocation to the larger population begins for a smaller $V$, $V=324$, above which the optimal solution is to maintain $V^1=324$ and devote the remainder of the vaccine to $ V^2$. 

This difference in strategy can be attributed to the fact that in the stochastic model, the gain $G$ begins to decline well before the onset of effective herd immunity.  This is evident in both the $N^1$ and $N^2$ solid curves of Figure 3A.  In contrast,  the gain in the deterministic model peaks very close to herd immunity at $V=400$, as the dashed curves of Figure 3A show.  Compared to the deterministic model, one can attribute this earlier decline in the average epidemic size $\langle E \rangle$ as being due to the probability of a terminal infection, which significantly lowers the average $\langle E \rangle$. More quantitatively, in Figure 3A, $V=324$ is the point at which the stochastic curve for $N^1$ crosses the initial value of curve $N^2$.

As $V$ increases further in Figure 5A, there is a sharp transition, indicating that if more vaccine exists than $V=474$ in the stochastic model or $V=657$ in the deterministic model, all vaccine should optimally be allocated to the large population.  This is the first switch noted earlier that occurs in both models.  As with the earlier transition, the first switch takes place at a smaller amount of vaccine $V$ in the stochastic model than in the deterministic model.  This sacrifices herd immunity that could have been achieved in the small population, in favor of relatively larger gains in protection that can be achieved with this level of vaccine in the large population.  Quantitatively it is clear from Figure 3A that beyond a certain amount of vaccine, the stochastic gain curve $G(V)$ begins to decline for $N^2$  while the stochastic curve for $N^1$ is still relatively large and constant.  Thus around this level of vaccine, all the available vaccine should optimally be switched into the larger population.  This same behavior is observed for the deterministic gain curves $\textbf{G}(V)$ for correspondingly larger values of $V$. 

Complete resource allocation to the large population continues until the large population achieves herd immunity, at which point a fraction of the vaccine is allocated to the smaller population. For the deterministic case, the optimal solution retains herd immunity for the large population, and increasingly allocates resources to the small population, until both populations achieve herd immunity. After that point, the optimal solution plateaus. For the deterministic model, the epidemic never progresses ($\textbf{I}(t) \leq \textbf{I}_0$). Because there is no cost for vaccination, remaining resources are allocated based solely on the relative population sizes (i.e. $1/3$ for the small population and $2/3$ for the large population). For the deterministic model, this corresponds to a situation with excess vaccine, since both populations are fully protected once each has sufficient resources to insure herd immunity. 

For the stochastic model, once there is sufficient vaccine to induce effective herd immunity in the large population, at around $V=660$, vaccine is once again allocated to the small population. However, unlike the deterministic case, for the stochastic model, there is a second abrupt shift in resources around $V=780$, resulting in a cusp in the optimal $V^1/V$, with the optimal solution approaching the final population based plateau value $V^1/V=1/3$ from above.  

This is due to the fact that in the stochastic model, for large $r_0$, the smaller population $N^1$ has a greater peak in gain.  Thus if there is enough vaccine available, there is a benefit to removing some vaccine from the large population in order to take advantage of the higher gain in the smaller population.  This second switch does not occur in the deterministic model because the opposite is true, the peak of deterministic curve $N^2$ for the larger population is always higher than the peak of deterministic curve $N^1$ for the smaller population in Figure 3A.  

For the stochastic model, the gain curves of Figure 3B can also be used to explain the absence of switching behavior for $r_0=2$, which is  generally observed for lower values of $r_0$ ($1<r_0\lesssim 2.9$).  A significant difference in this case is that the gain decreases continuously.  Due to the absence of peaks in the gain curve, the second switch observed for the large $r_0$ stochastic model, does not occur for small values of $r_0$.  The absence of the first switch is more subtle and depends on more than just the presence of a peak which exists when $r_0\gtrsim2.5$ as discussed previously. For the first switch to occur, the peak of the curve $G(V)$ must be large enough to offset the declines in the gain that first population $N^1$ exhibits.  Hence the first switch takes place for a more restrictive set of $r_0$, and only when the peak in the gain is sufficiently large ($r_0\gtrsim 2.9$).

In summary, the switching behavior of the optimal vaccination allocation are due, firstly, to the presence of peaks in the gain curves, and secondly, due to the relative heights of these peaks.  This explains why in the stochastic model, switching occurs only for large values of $r_0$, while in the deterministic model, it occurs for all values of $r_0>1$.  Fundamentally, this difference arises from the bimodal nature of the epidemic size distribution $P(E)$. 

\vspace{.35cm}
\noindent\textbf{The Range of Outcomes:} The differences in the optimal vaccination protocols between the stochastic and deterministic models can lead to substantial differences in the observed outcomes. The optimal protocols 
for the stochastic and deterministic models coincide for small quantities of vaccine, where in both cases it is optimal to allocate all vaccines to the smaller population. The optimal solutions also coincide in the limit of large quantities of vaccine, where it is optimal to allocate vaccines in proportion to the population size.   
The range of possible outcomes for different vaccination strategies indicates that optimization is most important when intermediate amounts of vaccine are available.  One way of understanding the potential impact associated with the optimal stochastic and deterministic protocols is by comparing their projected outcomes when applied to the presumably more realistic stochastic SIR model.  In this scenario, there is considerable difference between the best and worst possible outcomes and a significant but smaller difference between the stochastic and deterministic optimal solutions.

\begin{figure}[!htb]
\centering
\includegraphics[width=\linewidth,draft=false]{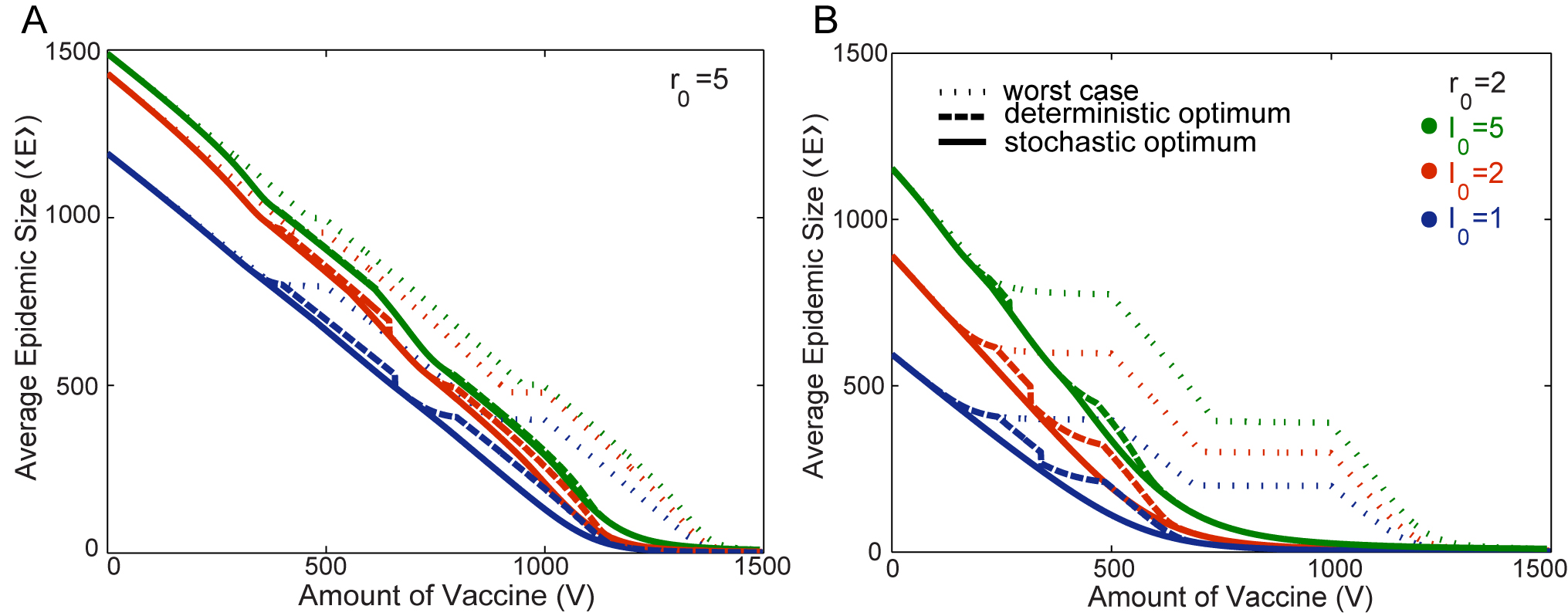}
\caption{\textbf{Comparison of Average Epidemic Size for Optimal, Deterministic Optimal, and Worst Case solutions:}  These figures illustrate the average epidemic size $\langle E \rangle$ for three different protocols: the stochastic optimum (minimizes $\langle E \rangle$), the deterministic optimum, and the worst case scenario (maximizes $\langle E \rangle$).  The three different colors correspond to different number of initial infected individuals $I_0$.  Results are obtained for two populations with size $N^1=500$ individuals and $N^2=1000$ individuals. Results are shown for $r_0=5$ in Figure A, and $r_0=2$ in Figure B. Other than the value of $r_0$, the color and line style legends in B apply to both graphs.}
\end{figure}

The dashed gold lines of Figures 5A and 5B illustrate the difference in the resultant average epidemic size $\langle E \rangle$ between the stochastic and deterministic optimal protocols.  Both protocols are the same in resource rich and resource poor regimes, and hence yield identical results.  Figures 6A and 6B illustrate $\langle E \rangle$ for both the stochastic and deterministic optimal protocols as well as the worst case allocation.

We define the \lq\lq worst case" protocol as that which maximizes $\langle E \rangle$ within the range of allowed allocations illustrated in Figure 4A and 4B.  Together the stochastic optimal solution and worst case allocation define the possible range of $\langle E \rangle$ at a given value of $V$.   Figure 6 shows that the difference between the outcome of the worst case protocol, and either the optimal stochastic and deterministic protocols, is substantially larger than the difference between the stochastic and deterministic cases.  This is particularly pronounced for smaller values of $r_0$, i.e. $r_0=2$. The worst case protocol would involve continuing to place vaccine in a population even after it is near or has reached herd immunity.  This is represented by the plateaus where the average epidemic size $\langle E \rangle$ is not significantly lowered by further vaccinating members of the population.  Deterministically, this is evident from the fact that $d\textbf{I}/dt<0$ as soon as the herd immunity threshold has been reached. The deterministic herd immunity threshold, serves as an approximate guide for when to stop vaccinating even in the stochastic case.

The differences between the stochastic and deterministic protocols have a complex $r_0$ and $I_0$ dependence.   The effect on the difference in $\langle E \rangle$ between the stochastic and deterministic models that is caused by increasing $I_0$  is different for small compared to large reproductive numbers $r_0$.  With a small reproductive number, e.g., $r_0=2$, the difference in the average epidemic size between the stochastic and deterministic optimal protocols is largest at an intermediate value of $I_0$, $I_0^1=I_0^2=2$ for the case illustrated in Figure 6B.   In contrast for large reproductive number, e.g. $r_0=5$, Figure 6A illustrates that the difference in the average epidemic size between the stochastic and deterministic optimal protocols is maximized for $I_0^1=I_0^2=1$ and decreases steadily as $I_0$ is increased.


\newpage
\subsection*{Alternate Cost Functions}

So far, we have defined the optimal allocation as that which minimizes the average epidemic size  $\langle E \rangle$, a quantity that contains contributions from both terminal infections and large-scale epidemics, but is not necessarily representative of any specific epidemic size that is likely to be observed because of the gap in the size distribution $P(E)$ (Figure 1A).  Choosing to minimize the deterministic result, which is the same as the average large-scale epidemic size \textbf{E}, might potentially be viewed as a conservative approach that safeguards against the case in which both populations will experience large-scale epidemics.

Other criteria for optimization may be considered within this framework.  For example, minimizing the maximum size rather than the average size could be the target for optimization.  To address this, in the same scenario of two non-interacting populations (e.g. two well-separated cities) with $N^1=500$ individuals and $N^2=1000$ individuals and both populations initiated with a single initial infected $I_0^1=I_0^2=1$, here we alternatively consider the probability that the epidemic is below some particular threshold tolerance size $E_{\mathrm{max}}$. 


\begin{figure}[!htb]
\centering
\includegraphics[width=\linewidth,draft=false]{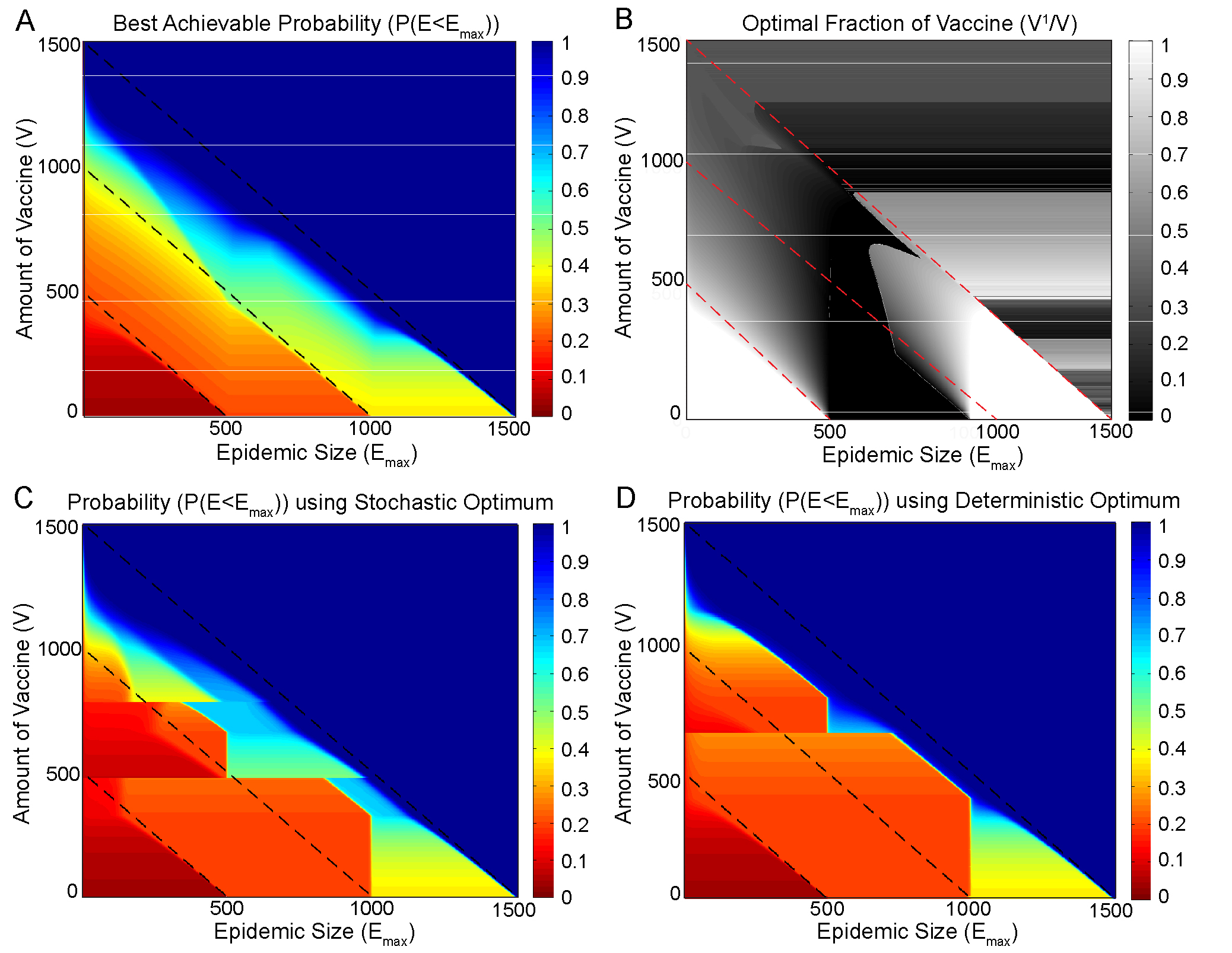}
\caption{\textbf{Optimizing the probability of having an epidemic less than a given size:}  Figure A shows in color, the optimal (largest) probability of having an epidemic less than some given size (x-axis), given some amount of vaccine (y-axis).  Figure 7B shows the fraction of vaccine $V^1/V$ in the smaller population that corresponds to optimal probability shown in Figure A.  Figures C and D show the same probabilities for, respectively, the deterministic and the stochastic solutions.   The dashed black lines can be used as reference points for comparing different figures.  Results are obtained for two populations with size $N^1=500$ individuals and $N^2=1000$ individuals, and both populations are initiated with a single infected individual $I_0^1=I_0^2=1$. }
\end{figure}

A policy maker may be interested in how much vaccine $V$ would be necessary and how it must be allocated between two populations in order to keep the total epidemic below some size $E_{\mathrm{max}}$.  We compute the best achievable probability of having an epidemic below a given size $E_{\mathrm{max}}$ given a total amount of vaccine $V$.  The results are shown in Figure 7A.  

The sharp color contrast of the diagonal bands in Figure 7A are associated with step-like changes in probability, arising from the bimodal nature of the epidemic size distributions $P(E)$. Because there is
very little probability for an event in the size range between the large-scale epidemic and the terminal infection peaks, when the threshold $E_{\mathrm{max}}$ passes through the large-scale epidemic size (which depends on the vaccine allocation) in the small population, the large population, or the sum of the two, nearly discrete steps in probability are observed.

The allocation that maximizes this probability is shown in Figure 7B, and is a function of both the amount of vaccine $V$ and also $E_{\mathrm{max}}$.  Unlike our previous optimization based on expected size (where the corresponding plot depends only on $V$), here the solution is extremely complex, switching discontinuously and frequently depending on both $V$ and $E_{\mathrm{max}}$, as indicated by sharp grey scale contrasts reflecting boundaries between high and low allocations to the small population.  In the resource poor regime (small $V$, corresponding to the lower horizontal boundary of the color plot) the solution switches from full allocation to the small population, to full allocation to the large population, back to full allocation to the small population. The lower left white triangle in Figure 7B corresponds to the situation with few resources, and minimal tolerance for the epidemic size. As in the previous stochastic and deterministic solutions aimed at minimizing the average epidemic size, here the optimal solution allocates all resources to the smaller population.
In the $E_{\textrm{max}}$ dependent resource rich regime, corresponding to points above the highest diagonal, the maximum achievable probability in Figure 7A is near unity, and the optimal allocation simplifies to depend only on $V$ (corresponding to horizontal bands in Figure 7B). 
However, in intermediate cases, where tradeoffs are most critical, the structure of the resulting solution is much too subtle to be realistically implemented for real populations given a limited amount of vaccine $V$.

For comparison, we evaluate the corresponding probabilities based on our previous stochastic and deterministic optimal protocols.  While both solutions are suboptimal for this alternative criterion, the stochastic solution comes close to the optimal case.  Figure 7C shows this result for the stochastic optimal solution, which replicates much of the green and blue high probability regions above the intermediate reference line.  It does a suboptimal job for relatively smaller epidemics in the regions where the amount of vaccine ranges from $V=400$ to $V=1000$.

Figure 7D illustrates the corresponding results when the optimal deterministic protocol is applied.  In maximizing $P(E<E_{\mathrm{max}})$,  the deterministic protocol  underperforms compared to the protocols of both Figure 7A and 7C.  Comparatively, the deterministic protocol minimizes the area of the high probability (blue) regions.  It  does slightly better than the stochastic optimum in roughly the same regions where the stochastic optimum fails compared to the best possible result, from about $V=400$ to $V=750$. 

This shows that the situation does indeed become more complicated when one looks beyond optimizing the average epidemic size  $\langle E \rangle$.  If the goal is to keep the epidemic below some size, given some amount of vaccine, there are indeed regions where the deterministically optimal solution will yield slightly better results.  Most of the time however, optimizing the average stochastic epidemic size gives a result closer to the best possible one of Figure 7A.  These figures thus indicate that the average epidemic size is a potentially useful metric for gauging the effects of stochasticity and will most of the time yield a solution that is preferable to the deterministic optimum.

\newpage
\section*{Discussion}

This paper illustrates the viability and power of developing the exact numerical solution of the master equations, done here for the stochastic SIR model. We generalized the method developed by Jenkinson and Goutsias~\cite{jenkinson} to obtain even greater numerical efficiency.  Instead of computing the probability of the system making a certain number and type of transitions between states, we directly compute the probability of the system residing in each state.  The advantage of this latter method is that it makes it easy to identify and eliminate excess states which are included by construction in the original method.  Even more significantly, our work and that of Jenkinson and Goutsias~\cite{jenkinson} provide proof of concept for obtaining accurate, exact solutions for SIR-type models, rather than relying on sampling methods~\cite{petrovic} or approximations to the master equation~\cite{keeling3}~\cite{munsky}.

Our analysis focuses on the fundamental tradeoff involves allocation of vaccine between two non-interacting communities of different size. Our procedure involved three steps. First, for each population we separately calculate the probability distribution of epidemic sizes for a given amount of vaccine. Second, we evaluate the expected epidemic size as a function of the amount of vaccine in each population. Third, we impose a constraint on the total amount of vaccine to distribute between the two populations, and determine the optimal allocation which minimizes the expected combined epidemic size of the two populations. 

We obtain several results that serve to elaborate and refine principles first identified by Keeling and Shattock~\cite{keeling2}, who considered the corresponding tradeoff in the context of the deterministic SIR model. Where the deterministic SIR model predicts a definite epidemic size for any given set of parameters, the stochastic SIR model produces a distribution, the characteristics of which significantly impact protocols for optimal allocation of vaccine. Under conditions that promote spread of the epidemic (i.e., the reproductive number $r_0>1$), the distribution of epidemic sizes obtained from the stochastic SIR model is bimodal~\cite{gordillo} in the limit of large population sizes, consisting of a peak describing terminal infections, that fail to propagate significantly in the population, and a peak describing large-scale epidemics, which have a mean size well approximated by the deterministic size. For finite population sizes, the distinction between terminal infections and large-scale epidemics vanishes at a value of $r_0$ that approaches unity as $N\to\infty$. 

Both the possibility of a terminal infection and the width of the distribution of the large-scale epidemic sizes contribute significantly to differences in the optimal allocation of vaccine for the stochastic model compared to the deterministic case. The differences are most significant for intermediate ranges of vaccine. In contrast, for both the stochastic and deterministic cases, when vaccine is severely limited or abundant, there is little or no difference in the optimal allocation of vaccine between the two models.

Differences in optimal allocations are amplified for intermediate amounts of vaccine because of the strong switching behavior of the optimal strategy. This switching can arise in both the stochastic and deterministic models, but at different points quantitatively, and is not always observed in the stochastic case. If the deterministic protocol is applied to the more realistic stochastic description of the epidemic evolution in the two populations, the performance is suboptimal, leading to a greater average epidemic size than would occur using the stochastic protocol. The difference is most significant for smaller values of $r_0$ where there is the most significant probability of a terminal infection.  The dependence on $I_0$, the number of infected individual, is more complex and depends on $r_0$, but in the limit where both $r_0$ and $I_0$ are large, the results converge to those of the deterministic SIR model. In the absence of vaccine, these quantities both increase the relative weight in the peak describing terminal infections. 

Keeling and Shattock~\cite{keeling2} attribute the switching behavior to the property of herd immunity, which occurs when the amount of vaccine is sufficient to prevent the epidemic from spreading significantly in the population. Herd immunity occurs in the deterministic SIR model when the initial effective growth rate of the number of infected individuals in the population becomes less than unity~\cite{jong}. While the optimal deterministic solution approximately distributes vaccine in a manner that achieves herd immunity in the largest possible population, this is not exactly the case. More precisely, the sharp transitions in both the deterministic and stochastic models arise from optimizing the overall impact of the vaccine in reducing the joint epidemic size, which we attribute to maximizing the overall gain. In the deterministic model, the impact of vaccine on epidemic size reduction is maximized as herd immunity is approached. For the stochastic model, the maximal impact typically occurs earlier, and in some cases, there is no sharp, intermediate transition.  

Policies involving strong switching may be difficult to implement publically, as one community could be reluctant to voluntarily sacrifice their entire vaccine allocation to another community in favor of a reduction in the overall epidemic size. In contrast, allocations in proportion to population size are likely to be less controversial to implement. Our observation that the stochastic model exhibits less dramatic switching for smaller values of $r_0$ suggests that such a policy could be justified as optimal in some cases. Furthermore, the reduction in magnitude of the gain peaks for the stochastic model in Figure 3, compared to the deterministic case, indicate that the overall magnitude of the benefit 
(as measured by reduction of the epidemic size), is less sensitive to the precise details of the allocation in the stochastic model than it is in the corresponding deterministic case. 

Interestingly, our analysis reveals that compared to the deterministic protocol, the stochastic protocol that minimizes the expected epidemic size, also overall better approximates an alternative target based on specifying a maximum tolerance (or threshold) for the overall epidemic size. This result is somewhat surprising. One might have expected the deterministic model to be more accurate in this case, because it predicts a large-scale epidemic whenever $r_0>1$, and as such might have captured a threshold criterion more accurately. The fact that the stochastic protocol continues to outperform the deterministic counterpart provides additional impetus to include the more complete and accurate stochastic dynamics of epidemic evolution in further studies. 

This paper isolates the tradeoff in vaccination allocation between two non-interacting populations, prior to the onset of widespread disease, in order to illustrate the significance of the full stochastic solution compared to deterministic case.   Our analysis relies on some strong assumptions, particularly the assumption of non-interacting populations. The extreme switching behavior in the deterministic case results from this non-interaction.  
It is less clear what the optimal policy will be for the case of weakly interacting populations.  One might speculate that the presence of even a modest amount of interaction yields dynamics of a single population, particularly in the deterministic case. 

Our conceptual framework and methods can potentially be generalized to include increasingly realistic situations, including interacting populations and real time allocation of vaccine as the epidemic evolves. 
In these scenarios, we anticipate detailed monitoring of stochastic effects, as well as incorporation of delays associated with transportation and the onset of immunity, will play a critical role in determining the optimal dynamic protocol, and we expect that the critical differences between the stochastic and deterministic SIR models illustrated here will have an increasingly significant impact in identifying protocols that aid in minimizing the overall epidemic size.  

The hope is that the systematic study of such tradeoffs will shed light on the development of effective policies.
For example, in the case of epidemic outbreak in a localized geographic region, government officials might have to decide whether to allocate scare vaccination doses exclusively to that region or to allocate the vaccine proportionately for population as a whole.  In situations where vaccine doses have been prepositioned geographically, the question of ``giving away'' vaccines from one region to another will be of intense debate. Thus, issues of fairness will complicate decisions even more.  Identifying policies that are close-to-optimal and can actually be implemented is an important topic for future research.

\section*{Acknowledgments}
The authors thank Charles Lieou and Kimberly Schlesinger for helpful discussions and feedback. This work was supported by an Office of Naval Research MURI Grant No. DMR0606092, the David and Lucile Packard Foundation, the Institute for Collaborative Biotechnologies through contract no. W911NF-09-D-0001 from the U.S. Army Research Office, and the Stansberry Fellowship through the CCS SURF foundation.
\bibliography{mybib}{}

\end{document}